\documentclass[
 reprint,
 amsmath,amssymb,
 aps,
]{revtex4-2}

\usepackage{graphicx}          
\usepackage{dcolumn}           
\usepackage{bm}                
\usepackage{hyperref}          
\usepackage[mathlines]{lineno} 

\begin{document}

\preprint{APS/123-QED}

\title{Multi-Tone Microwave Frequency Locking to a Noisy Cavity via Real-Time Feedback}

\author{J.P. van Soest}
\email{j.vansoest-1@tudelft.nl}
\affiliation{Kavli Institute of NanoScience, Delft University of Technology, PO Box 5046, 2600 GA Delft, Netherlands}
\author{C.A. Potts}
\affiliation{Kavli Institute of NanoScience, Delft University of Technology, PO Box 5046, 2600 GA Delft, Netherlands}
\author{S. Peiter}
\affiliation{Kavli Institute of NanoScience, Delft University of Technology, PO Box 5046, 2600 GA Delft, Netherlands}
\author{A. Sanz Mora}
\affiliation{Kavli Institute of NanoScience, Delft University of Technology, PO Box 5046, 2600 GA Delft, Netherlands}
\author{G.A. Steele}
\email{g.a.steele@tudelft.nl}
\affiliation{Kavli Institute of NanoScience, Delft University of Technology, PO Box 5046, 2600 GA Delft, Netherlands}
\date{\today}

\begin{abstract}
Microwave cavities are commonly used in many experiments, including optomechanics, magnetic field sensing, magnomechanics and circuit quantum electrodynamics. Noise, such as variations in magnetic field or mechanical vibrations, can cause fluctuations of the cavity's natural frequency, creating challenges in operating them in experiments. To overcome these challenges, we demonstrate a dynamic feedback system implemented by the locking of a microwave drive to the noisy cavity. A homodyne interferometer scheme monitors the cavity resonance fluctuations due to low-frequency noise, which is mitigated by frequency-modulating the microwave generator. The feedback has a bandwidth of $400$ Hz, with a reduction of cavity fluctuations by $85\%$ integrating up to a bandwidth of $2$ kHz. Moreover, the cavity resonance frequency fluctuations are reduced by $73\%$. This scheme can be scaled to enable multi-tone experiments locked to the same feedback signal. As a demonstration, we apply the feedback to an optomechanical experiment and implement a cavity-locked, multi-tone mechanical measurement. As low-frequency cavity frequency noise can be a limiting factor in many experiments, the multi-tone microwave locking technique presented here is expected to be relevant for a wide range of microwave cavity experiments.
\end{abstract}

\maketitle

\section{Introduction}\label{Introduction}
Low-frequency noise is often the limiting factor in mechanical and magnetic experiments \cite{Oosterkamp2014,niepce2021stability,Learn2022,Brock2020,Muller2015,vadakkumbatt2021prototype,Bouwmeester2017,Regal2008,clark2018cryogenic}, arising, for example, from variations in magnetic field or vibrations in cryogenic experiments. These sources of noise can cause fluctuations in a microwave cavity's resonance. There are two common ways to counteract this noise: the first is to dampen the noise before it reaches the device under test. While a mass-on-a-spring low-pass filter can suppress high-frequency vibrations for mechanical experiments \cite{Oosterkamp2019,lee2018vibration}, the lowest frequencies can often introduce considerable noise. In this sense, the mechanical filter is not optimal: it is effective at blocking high frequencies but not low frequencies. The second method actively counters noise by applying dynamic feedback to the system \cite{Kanhirathingal2022,vepsalainen2022improving}. As a feedback system is limited by its feedback bandwidth and response time, it acts as a high-pass filter: it cannot react at high frequencies but is very effective at low frequencies. Combining these two complementary techniques, therefore, gives an opportunity for noise reduction over the entire spectrum.\\
\indent In this work, we present an active feedback system based on the cavity stabilization method introduced by Pound in Ref.~\cite{Pound1946}. Here, the inverse of the Pound-Drever-Hall technique \cite{Drever1983, Ortiz1983, Black2001, Vahala2005} is applied using active feedback to adjust a microwave generator frequency, locking it to a cavity whose resonance frequency fluctuates in time. This scheme not only locks a single microwave signal to the cavity, but also enables the active locking of multiple tones using the same feedback signal, demonstrated in Section.~\ref{Multi-Tone_Locking}.

\begin{figure}[b!]
    \centering
    \includegraphics{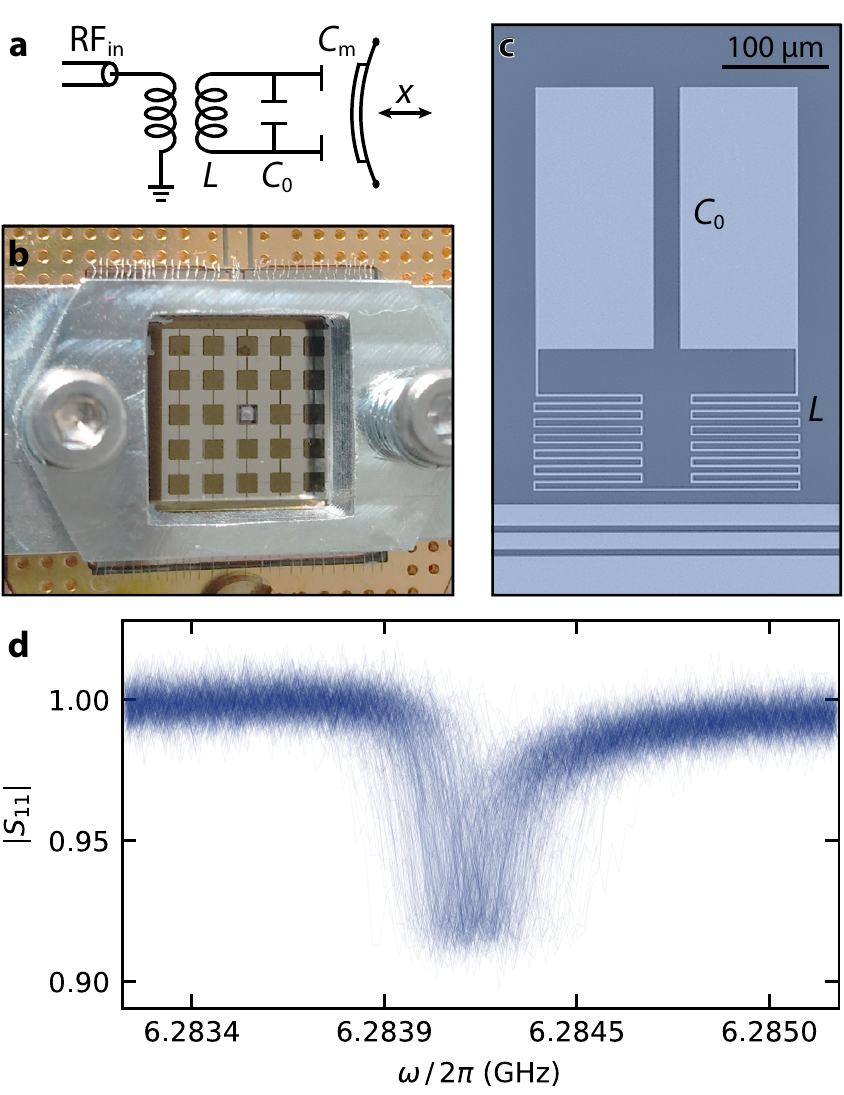}
    \caption{A microwave cavity with resonance frequency fluctuations. (a) Lumped element circuit schematic of the device under test. (b) Optical image of the flip-chip device. The metallized membrane can be seen as a white square in the phononic shield. (c) Optical micrograph of the RF circuit. The meander inductor $L$ and capacitor $C_{\rm 0}$ form the microwave cavity. (d) $500$ individual normalized $240$ \textmu s VNA traces show a distribution of the cavity resonance frequency $\omega_{\rm c}$.}
    \label{Figure1}
\end{figure}

\begin{figure*}[t!]
    \centering
    \includegraphics{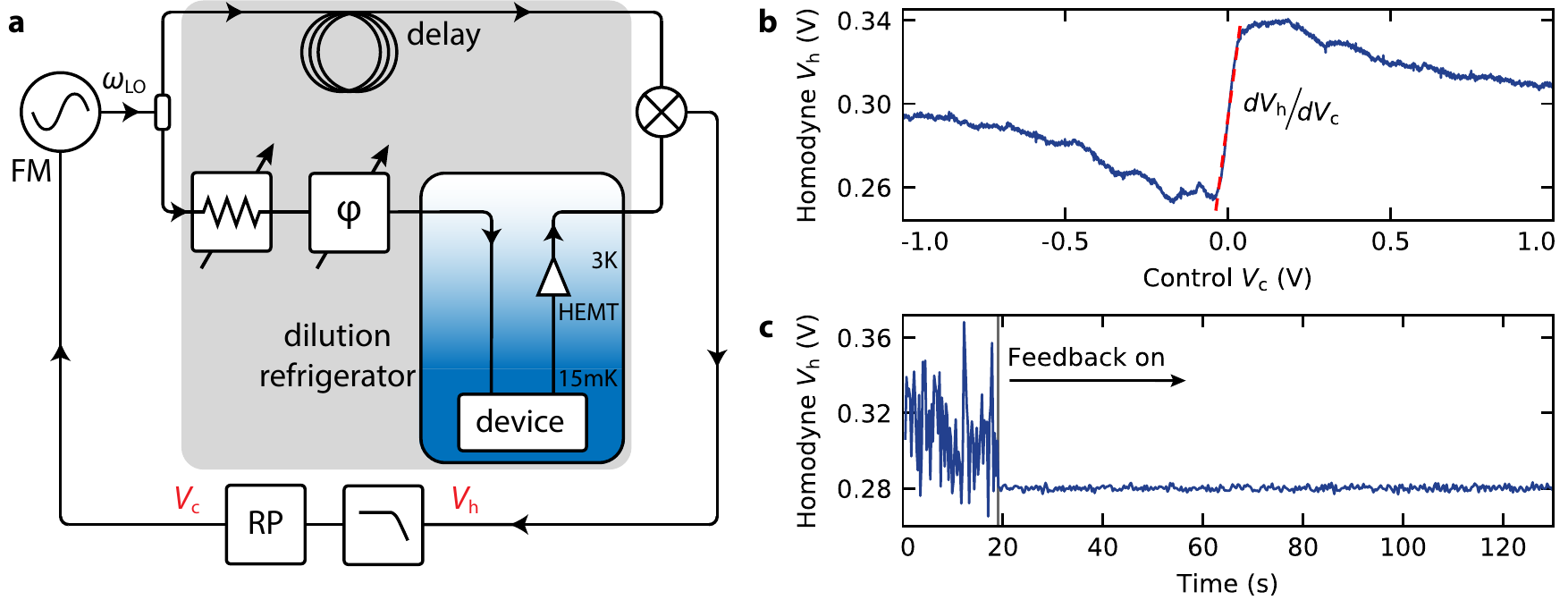}
    \caption{The feedback system is used to reduce the cavity resonance frequency fluctuations. (a) Schematic of the feedback circuit, consisting of a balanced microwave homodyne interferometer, of which one path goes to the device under test. The reference arm is path length matched, downconverting the signal at a mixer. The resulting voltage $V_{\rm h}$ is led to a Red Pitaya (RP), which creates a control voltage $V_{\rm c}$ to modulate the frequency of the local oscillator $\omega_{\rm LO}$. (b) Homodyne response as the control voltage $V_{\rm c}$ sweeps the local oscillator frequency over the cavity resonance. Within the linewidth of the cavity, the response is linear. Low-frequency noise can be seen outside the linear regime. (c) Time trace of the homodyne voltage $V_{\rm h}$. The feedback is turned on at $t = 19$ seconds.}
    \label{Figure2}
\end{figure*}

\indent Since any source of low-frequency cavity noise can be mitigated, this technique has applications in various fields. It can be used to modulate magnetic flux to stabilize SQUID cavities \cite{vepsalainen2022improving,bothner2021photon}, and magnomechanical experiments \cite{potts2021dynamical,potts2022dynamical,zhang2016cavity,shen2022mechanical}, which are also known to suffer from magnetic field noise. As we demonstrate in Section.~\ref{Optomechanical_Experimental_Validation}, our technique can be directly applied to an optomechanical system experiencing mechanical noise \cite{aspelmeyer2014cavity}. As identified in previous work, cavity frequency noise from vibrations can be a limiting factor in ground state cooling of mechanical resonators \cite{Peiter2022}.

\section{Experimental Setup}\label{Experimental_Setup}
To demonstrate the feedback system, we apply it to a radio frequency (RF) circuit in an optomechanical flip-chip setup, similar to Ref.~\cite{noguchi2016ground,brubaker2022optomechanical}. The microwave cavity consists of a meander inductor $L$ and a pair of capacitor plates $C_{\rm 0}$ etched into a NbTiN film, onto which a mechanically compliant capacitor $C_{\rm m}$ is positioned\break $\sim\!1$ micron above. A circuit schematic of the device and optical images are shown in Fig.~\ref{Figure1}(a-c). The top capacitor plate consists of aluminum deposited on a thin silicon nitride membrane, embedded within a patterned silicon substrate etched to form a phononic shield; for example, see Ref.~\cite{brubaker2022optomechanical,delaney2022superconducting}. The primary purpose of the phononic crystal structure is to reduce unwanted mechanical noise at $1$ MHz and to suppress the radiative loss of mechanical energy through the substrate, as shown in Ref.~\cite{Peiter2022}. However, the phononic crystal has the unwanted side effect of increasing the mechanical susceptibility of the top chip to low-frequency ($\sim$kHz) vibrations in the setup \cite{fedorov2020thermal}.\\
\indent The resonance frequency of the flip-chip cavity is given by $\omega_{\rm c}(t) = [L (C_{\rm 0} + C_{\rm m}(t))]^{-1/2} \approx 2\pi \times 6.28$ GHz. Vibrations drive low-frequency modes of the top chip, resulting in a time-dependent capacitance $C_{\rm m}(t)$. Therefore, these low-frequency mechanical vibrations of the top chip modulate the cavity resonance frequency $\omega_{\rm c}$. The fluctuations in the cavity resonance frequency can be seen in Fig.~\ref{Figure1}(d), where many repeated VNA sweeps demonstrate the noise imparted on the microwave resonator. These measurements were performed on the base plate of a dilution refrigerator held at $\sim\!20$ mK. The pulse tube cooler is the primary source of vibrations within a dry dilution refrigerator; therefore, the pulse tube was switched off during measurements to reduce vibrations within the experimental setup. However, we still observed that the resonance frequency fluctuations are on the order of the linewidth of the cavity. To compensate for these fluctuations, we implemented an active feedback system to track the cavity's resonance frequency. 

\section{Feedback System}\label{Feedback_System}
\begin{figure*}[t!]
    \centering
    \includegraphics{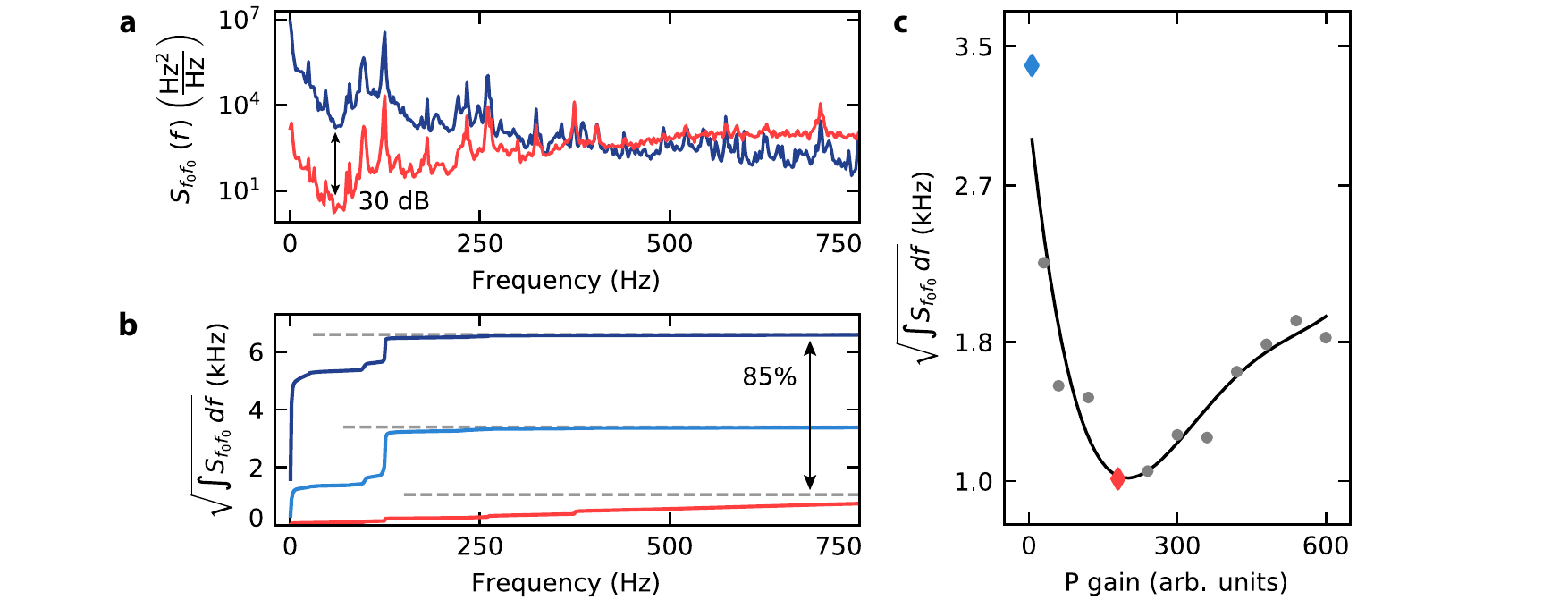}
    \caption{Noise reduction of the homodyne voltage $V_{\rm h}$ depending on the feedback gain. (a) The power spectral density $S_{f_{\rm 0}f_{\rm 0}}$ of the homodyne voltage $V_{\rm h}$ with the feedback off (dark blue) and on (red). The unlocked data has been calibrated as described in Appendix~\ref{SpeactralCal}. Noise is reduced up to $30$ dB over a bandwidth of $400$ Hz. (b) Integrated power spectral density for unlocked (dark blue) and locked cavity for two PI gains (light blue and red). Asymptotes up to $2$ kHz (dashed gray) show a noise reduction of $85\%$. (c) Integrated power spectral density up to $2$ kHz for different gains of the feedback loop, using the same color coding. The solid black line is a polynomial fit to guide the eye.}
    \label{Figure3}
\end{figure*}

To compensate for the fluctuating cavity frequency, we have developed a feedback system similar to the Pound-Drever-Hall technique \cite{Drever1983, Ortiz1983, Black2001, Vahala2005}. While the Pound-Drever-Hall technique is usually used to stabilize a noisy laser by locking it to a stable optical cavity, here we apply its inverse role: we lock a stable generator to the resonance frequency of a noisy cavity. In our system, the microwave frequency of the drive tone is locked to the fluctuating resonance frequency of the cavity. Locking the drive tone is accomplished by producing an error signal depending on the time-dependent resonance frequency of the cavity. The error signal is then used in an active feedback loop to frequency-modulate (FM) the microwave generator's output, producing a drive tone that tracks the shifting cavity resonance frequency. A schematic diagram of the feedback system is shown in Fig.~\ref{Figure2}(a).\\
The scheme starts by splitting the local oscillator signal into two paths: one of which goes directly to the microwave cavity, then it is directed into the RF port of a mixer. The other path goes through a path length matched reference arm to the LO port of the same mixer, making the system a balanced homodyne microwave interferometer. Small fluctuations in the resonance frequency of the microwave cavity result in a linear response at the output of the mixer, as seen in Fig.~\ref{Figure2}(b). This downconverted homodyne signal, which we call the homodyne voltage $V_{\rm h}$, depends on the detuning of the cavity resonance frequency relative to the current microwave drive frequency. When the system is unlocked, the homodyne voltage fluctuations follow the fluctuations in the cavity's resonance frequency. In this experiment, the cavity fluctuations are larger than the linear region of the homodyne downconversion. Thus, the local oscillator frequency is updated in real-time by modulation using a control voltage $V_{\rm c}$ to track the cavity resonance frequency.\\
\indent Sweeping the local oscillator's frequency $\omega_{\rm LO}$, a linear dependency between the control voltage $V_{\rm c}$ and the measured homodyne voltage $V_{\rm h}$ was observed around the cavity resonance, as seen in Fig.~\ref{Figure2}(b). Notice that if an unbalanced homodyne detection scheme were used, the path length difference would result in phase-winding fringes in the homodyne voltage signal, which could result in a loss of cavity lock. Thus, using a balanced homodyne scheme ensures a frequency-independent homodyne voltage signal regardless of the instantaneous local oscillator frequency, see Appendix~\ref{PhaseMatch}.\\
\indent To maintain constant power arriving at the LO port of the mixer while having dynamic control over the power incident on the microwave resonator under test, a variable attenuator was added to the device arm. Moreover, the phase difference between the local oscillator and device arm was set to $90^{\circ}$ out of phase, so the quadrature signal was measured. The homodyne voltage is then low-pass filtered and led to a proportional-integral (PI) controller programmed in the field-programmable gate array (FPGA) of a Red Pitaya \cite{RedPitaya}. The output control voltage of the PI controller $V_{\rm c}$ continuously modulates the local oscillator's frequency $\omega_{\rm LO}$ to remain resonant with the cavity frequency $\omega_{\rm c}$. Ideally, the homodyne voltage is zero at resonance; however, a slight DC offset was present in the experiment, which determined the setpoint in the PI controller.\\
\indent To optimize our feedback system, the proportional and integral gain of the PI controller needed to be tuned for optimal noise suppression. As the gain is increased, low-frequency noise is suppressed; however, a too-aggressive PI setting results in added noise at higher frequencies. The optimal value for all experimental implementations of our feedback system depends on the specific configuration; therefore, each system must be calibrated to find the optimal PI gains. Turning on the feedback system stabilizes the homodyne voltage, as can be seen in Fig.~\ref{Figure2}(c). It should be noted that the feedback did not lose the lock during the measurements over a time of approximately $7$ minutes (the period that the pulse tube could be turned off).

\section{Noise Reduction}\label{Noise_Reduction}
\begin{figure}[t!]
    \centering
    \includegraphics{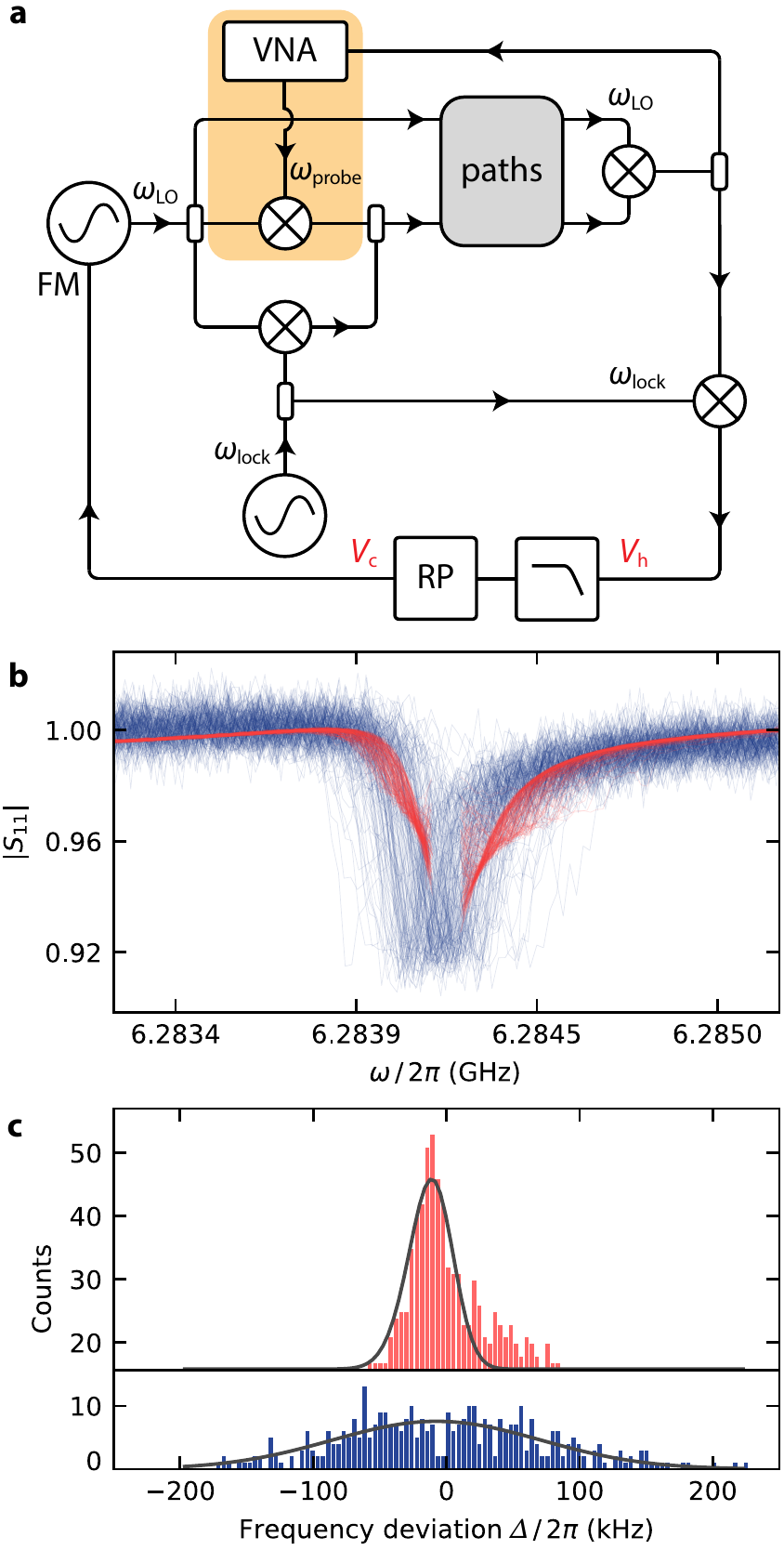}
    \caption{Characterization of the locked cavity response measured with the extended feedback circuit. (a) Schematic of the extended feedback circuit to lock multiple signals to the same feedback. The gray box corresponds with that in Fig.~\ref{Figure2}. The local oscillator is now up- and downconverted with a lock tone. The components in the orange box are included to add a probe tone to the input of the device. (b) Normalized VNA traces with the cavity unlocked (blue) and locked (red). Two separate measurements are performed for the locked cavity. (c) Histogram of the fitted resonance frequency of $376$ traces unlocked and locked, using the same color coding. The locked histogram has been shifted for clarity. The standard deviation of the locked cavity resonance frequency is $\sigma_{\rm locked} = 21$ kHz, which is a factor $4$ smaller than the unlocked $\sigma_{\rm unlocked} = 77$ kHz.}
    \label{Figure4}
\end{figure}

We can quantitatively analyze the suppressed frequency noise in the reference frame of the cavity by considering the power spectral density of frequency fluctuations $S_{f_{\rm 0}f_{\rm 0}}(f)$ (Hz$^2$/Hz) \cite{fedorov2020thermal}. The power spectral density can be calculated by taking the Fourier transform of the measured homodyne voltage $V_{\rm h}$ and is shown in Fig.~\ref{Figure3}(a). The power spectral density of the locked homodyne voltage $S_{V_{\rm h}V_{\rm h}}(f)$ (V$^2$/Hz) was calibrated using the slope of the homodyne voltage versus control voltage, shown in Fig.~\ref{Figure2}(b). However, the unlocked homodyne signal's power spectral density cannot be calibrated this way, since the region of linear response is narrower than the fluctuations of the cavity resonance frequency. Therefore, the detected homodyne signal does not contain information of large frequency fluctuations. An extra calibration has been applied to the unlocked spectra, as discussed in Appendix~\ref{SpeactralCal}.\\
\indent We find that when we turn the feedback on, even with a small gain, the large frequency shifts of the cavity are compensated for, and the cavity fluctuations remain within the linear regime of the homodyne signal. This indicates that low-frequency fluctuations dominate the cavity noise. We identified that for our system the optimal proportional and integral gain ratio is $\rm P \, \colon I = 6 \ \colon 10$. Figure~\ref{Figure3}(a) shows the power spectral density of the cavity unlocked and locked with the optimal feedback loop gain. It can be seen that for low frequencies, the noise is suppressed up to $30$ dB. The spectra cross around\break $400$ Hz, indicating the feedback bandwidth. Above this frequency, the system adds noise to the measurement, due to feedback of the added noise by the amplification chain used. At $2$ kHz and above, the two spectra overlap, suggesting no additional high-frequency noise is being added by our feedback system. Taking the square root of the integrated power spectral densities $S_{f_{\rm 0}f_{\rm 0}}(f)$, we obtain the root mean square (rms) value for the noise power in units of Hz. Even though there is added noise at intermediate frequencies, in Fig.~\ref{Figure3}(b) it can be seen that the total frequency noise is reduced. The integrated spectra show a noise reduction of $85\%$ at $2$ kHz, as indicated by the asymptotes shown in Fig.~\ref{Figure3}(b). Again, this suggests that low-frequency fluctuations were most detrimental in our system. Since the feedback loop acts as a high-pass filter, these low frequencies are compensated most efficiently.\\
\indent Finally, the total gain of the feedback loop was varied by increasing the PI gain while keeping their ratio the same. Increasing the gain initially results in additional noise suppression until the feedback increases the total noise power, as shown in Fig.~\ref{Figure3}(c). This provides an optimal feedback loop gain for this system. 

\section{Multi-Tone Locking}\label{Multi-Tone_Locking}
Thus far, all measurements were done using a single tone on cavity resonance. However, many experiments require multiple tones or, in fact, frequency sweeps. Since the local oscillator follows the cavity's resonance frequency, the feedback circuit can easily be scaled up by adding additional generators to the same feedback setup. For example, an extra stage of up- and downconversion was added to the circuit, as shown in Fig.~\ref{Figure4}(a). The gray box corresponds with that shown in Fig.~\ref{Figure2}(a). The local oscillator signal is now split into three paths: two paths in parallel are upconverted, to a lock signal $\omega_{\rm lock} = \omega_{\rm c} \: - \: \omega_{\rm LO}$ and a probe signal $\omega_{\rm probe}$ generated by a VNA; the third path goes to the reference arm as before. The signal coming from the cavity is first downconverted using $\omega_{\rm LO}$, producing two tones at $\omega_{\rm lock}$ and $\omega_{\rm probe}$. The signal at $\omega_{\rm probe}$ can be measured by the VNA as long as $\omega_{\rm lock}$ is not within the IF bandwidth of the measurement. The other path (at $\omega_{\rm lock}$) is downconverted a second time to DC, resulting, as before, in a voltage containing the frequencies of the cavity resonance fluctuations. This signal will also contain fluctuations at the frequency $\delta = \omega_{\rm probe} - \omega_{\rm lock}$. If $\delta$ is within the feedback bandwidth of the cavity lock, it will need to be filtered out not to modulate the local oscillator with this frequency. Again, the downconverted signal is then led to the Red Pitaya to generate an error signal to frequency modulate the local oscillator.\\
\indent Figure~\ref{Figure4}(b) shows VNA measurements with the feedback off and on, where the same number of traces were used. Note that these measurements were performed with an IFBW of $600$ and $1$ kHz, respectively, and with feedback gains $\rm P = 480$, $\rm I = 800$. The locked measurement was performed on the left and right sides of the cavity separately to avoid the lock tone entering the VNA measurement, as discussed above. However, if a measurement requires a VNA sweep on cavity resonance, it is also possible to lock off-resonant, see Appendix~\ref{PhaseMatch}.\\
\indent The individual traces have been fit using a Lorentzian function to determine the resonance frequency for each trace. The extracted resonance frequency of the unlocked and locked traces are compared in Fig.~\ref{Figure4}(c). The left and right datasets are fit separately, using a fixed asymmetry due to the Fano effect \cite{rieger2022fano} extracted from the unlocked measurement. Here ${\it \Delta}$ is the deviation from the average value of the fit resonance frequency. The fits give skewed Gaussian distributions with a standard deviation $\sigma_{\rm unlocked} = 77$ kHz and $\sigma_{\rm locked} = 21$ kHz, which is a decrease of $73\%$ for the locked measurement.

\section{Optomechanical Experimental Validation}\label{Optomechanical_Experimental_Validation}
\begin{figure}[t!]
    \centering
    \includegraphics{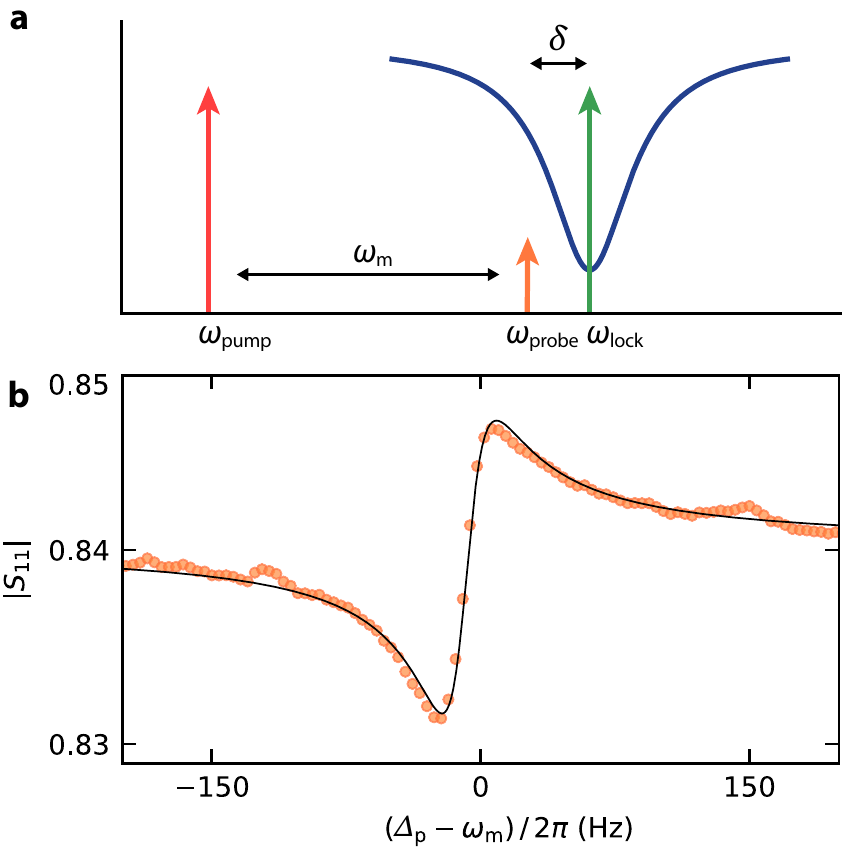}
    \caption{Demonstration of a cavity-locked multi-tone optomechanical measurement. (a) Illustration of the driving scheme. The upconverted lock tone is set on the cavity resonance. The pump tone is applied $\omega_{\rm m}$ away from the center of the probe tone of the VNA, which in turn is detuned by $\delta$ from resonance. (b) The mechanical resonance is visible as an OMIT peak, centered around ${\it \Delta}_{\rm p} = \omega_{\rm probe} - \omega_{\rm pump}$, when the local oscillator is locked to the cavity, with $\delta \approx 150$ kHz. Normalized data is shown in orange with a fit in black.}
    \label{Figure5}
\end{figure}

As a demonstration of the applicability of the feedback system, it is used in an optomechanical experiment \cite{aspelmeyer2014cavity}. The metal film of the capacitor $C_{\rm m}$ is located on a high-tensile stress silicon nitride membrane, parametrically coupling the membrane to the cavity \cite{noguchi2016ground}. The membrane is isolated from its surroundings as it is located in a phononic shield \cite{Peiter2022}. However, implementing this lattice structure also poses difficulties, as low-frequency modes are excited by vibrations in the dilution refrigerator. Although the energy of these lattice vibrations will mostly not be transferred to the vibrational mode of the membrane, it does mean that the membrane is located on a moving structure. As these modes alter the distance between the two chips by an amount larger than the amplitude of the vibrational mode of the membrane, this noise will change $C_{\rm m}$, and thus also $\omega_{\rm c}$, significantly.\\
\indent As introduced in Section \ref{Multi-Tone_Locking}, multiple microwave tones can be sent to the device while locked to the same local oscillator. Simply by multiplexing generators and mixers after the local oscillator, more drive frequencies can be added. Repeating the elements inside the orange box Fig.~\ref{Figure4}(a). offers continual scaling.\\
\indent Figure~\ref{Figure5}(a) shows the scheme for a multi-tone measurement. By applying a drive tone detuned from the red sideband of the cavity $\omega_{\rm pump} = \omega_{\rm c} -\, \omega_{\rm m} - \delta$ and probing close to resonance $\omega_{\rm probe} = \omega_{\rm c} - \delta$ the mechanical resonance appears from optomechanically induced transparency (OMIT) \cite{Kippenberg2010,safavi2011electromagnetically, Teufel2011}. While having the local oscillator locked, a measurement of the mechanical resonance is presented in Fig.~\ref{Figure5}(b). Note that the asymmetry of the OMIT peak is due to the detuning $\delta \approx 150$ kHz of the pump and probe with respect to the lock tone. However, it is equally possible to instead apply the detuning to the lock tone by adjusting the setpoint of the feedback circuit, see Appendix~\ref{PhaseMatch}.

\section{Conclusions}\label{Conclusions}
In conclusion, we have presented a dynamic feedback system locking multiple microwave tones to an unstable microwave cavity. A local oscillator's frequency was modulated to follow the cavity resonance frequency by calibrating the homodyne signal to a control voltage. The power spectral density of the homodyne voltage has been analyzed, revealing a noise reduction of $85\%$ at $2$ kHz. Optimization of the feedback loop and the PI controller resulted in an optimal response for our system with a feedback bandwidth of $400$ Hz. Furthermore, we extended the feedback system to enable multi-tone experiments. Additional microwave tones were implemented by upconversion with the same locked tone of the local oscillator. Moreover, VNA measurements probing the cavity showed that fluctuations in the cavity resonance frequency when locked had a standard deviation of $\sigma_{\rm locked} = 21$ kHz, as opposed to an unlocked measurement with $\sigma_{\rm unlocked} = 77$ kHz, which is an improvement of $73\%$.\\
\indent Finally, we demonstrated the applicability of the feedback system by performing an optomechanical experiment. We observed OMIT using another locked tone from a third generator without losing the cavity lock. Our feedback system provides continual scaling for adding tones to the same lock by multiplexing generators and mixers after the local oscillator. Besides optomechanics \cite{Teufel2016,Schwab2013}, the feedback system presented here has applications in various fields, as any source of low-frequency cavity noise can be reduced. For example, other potential applications include reducing TLS noise \cite{Lindstrom2011}, locking to coplanar waveguide resonators \cite{Gao2007}, SQUID cavities \cite{bothner2021photon}, gravitational wave detectors \cite{vadakkumbatt2021prototype}, axion dark-matter detectors \cite{khatiwada2021axion}, and magnomechanical experiments \cite{potts2021dynamical,potts2022dynamical,zhang2016cavity,shen2022mechanical}.

\begin{acknowledgments}
This work was supported by the European Union’s Horizon 2020 research and innovation programme under grant agreements 681476 - QOM3D and the Dutch Research Council (NWO) through a Vici award from the NWO Talent Programme. C.A. Potts acknowledges the support of the Natural Sciences and Engineering Research Council of Canada (NSERC). 
\end{acknowledgments}

\appendix

\section{Homodyne Phase Matching}\label{PhaseMatch}
\begin{figure}[t!]
    \centering
    \includegraphics{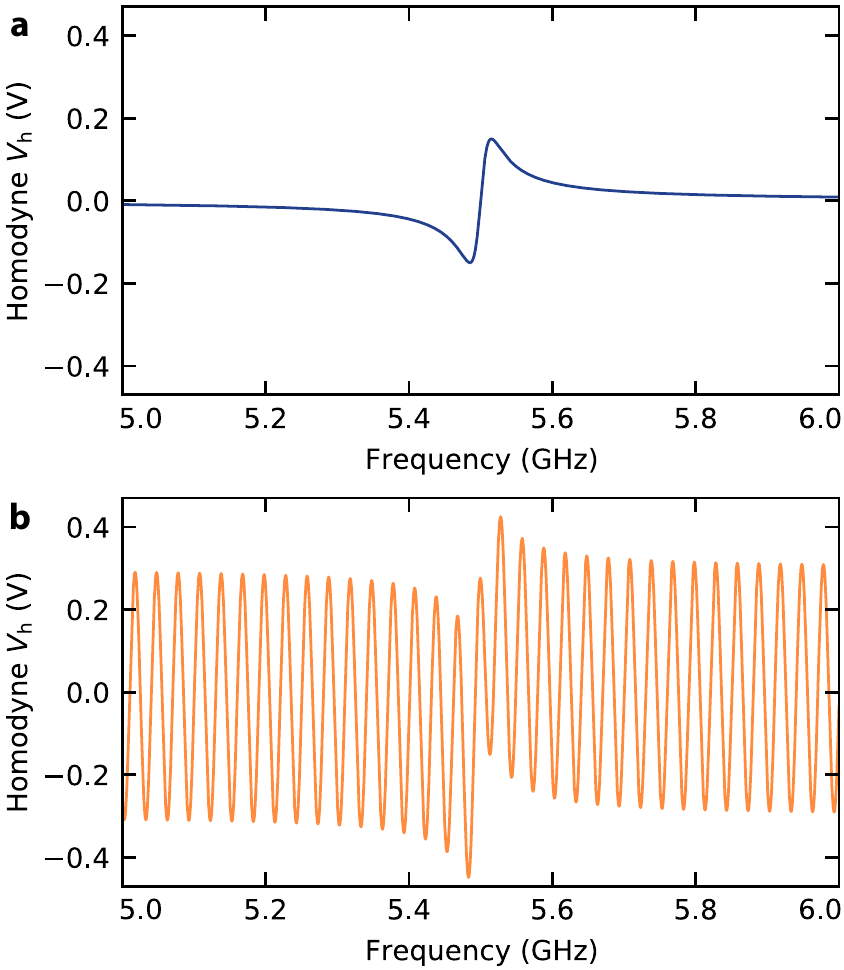}
    \caption{Homodyne response for different interferometer path lengths. (a) Calculations for a balanced interferometer. (b) Calculations for an unbalanced by ${\it \Delta L} = 10$ m interferometer.}
    \label{FigureSI1}
\end{figure}

\begin{figure}[t!]
    \centering
    \includegraphics{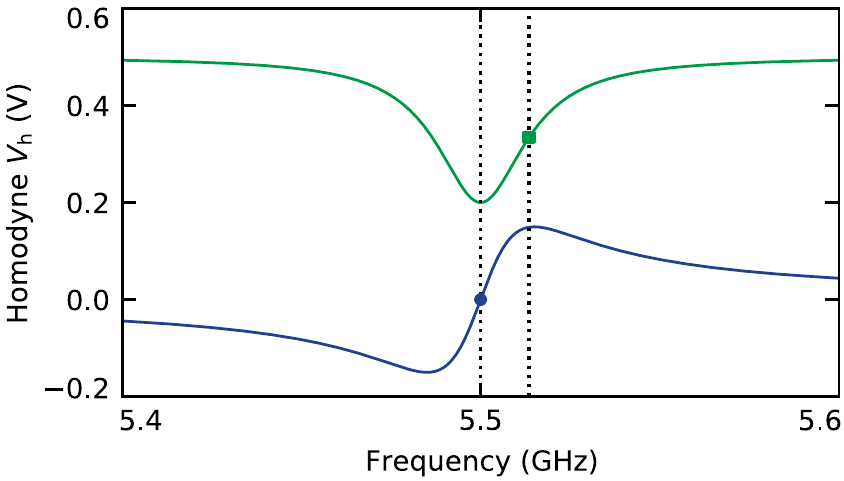}
    \caption{Calculated homodyne response for two quadratures, as defined by the phase difference of the two signals coming into the mixer. Locking on resonance can be achieved using the "phase" quadrature (blue) with a tone resonant with the nominal cavity frequency and a feedback setpoint indicated by the blue circle. Locking off resonance can be achieved by choosing the orthogonal "amplitude" quadrature (green) and a setpoint indicated by the green square.}
    \label{FigureSI2}
\end{figure}

It is worth discussing the effect of the balanced interferometer. As mentioned in Section \ref{Feedback_System}, the path length of the reference arm of the interferometer is matched to that of the signal arm. In contrast to the most common use of an interferometer, where the absolute phase difference of the two signals is not important, here we need to match the phases exactly. Because the noise in the system fluctuates the resonance frequency of the cavity, our feedback system continually changes the frequency of the drive tone and, thereby, its wavelength. Therefore, the signals arriving at the mixer will have a frequency-dependent phase difference unless the two path lengths are identical.\\
\indent Figure~\ref{FigureSI1} shows the calculated homodyne voltage $V_{\rm h}$ for a balanced and a ${\it \Delta L} = 10$ m unbalanced interferometer. In both cases, the incoming signals are set to be $90^{\circ}$ out of phase to measure the quadrature signal. It is clearly visible that in the unbalanced case, there is a sinusoidal profile superimposing the frequency response of the cavity, known as phase-winding fringes. The linear regime has decreased significantly, increasing the chance for the lock to fringe hop to the next slope. This would result in an unstable feedback signal.\\
\indent Moreover, it is possible to send the lock tone off-resonant from the cavity by adjusting the setpoint as long as the tone is within the linear regime of the homodyne response. However, one advantage of the homodyne voltage locking, in contrast to the conventional Pound-Drever-Hall technique \cite{Drever1983, Ortiz1983, Black2001, Vahala2005}, is that one can also choose to lock at other points in the cavity. Doing so will sacrifice some stability, which will reduce the quality of the lock, but choosing the right phase difference at the mixer allows one lock on the other quadrature, as demonstrated in Figure~\ref{FigureSI2}. The dashed lines show examples of setpoints that allow locking on- and off-resonant, while still remaining on an unidirectional slope, using the other quadrature. Note that moving the setpoint further from resonance can decrease the stability of the lock.

\section{Spectral Density Calibration}\label{SpeactralCal}
\begin{figure}[t!]
    \centering
    \includegraphics{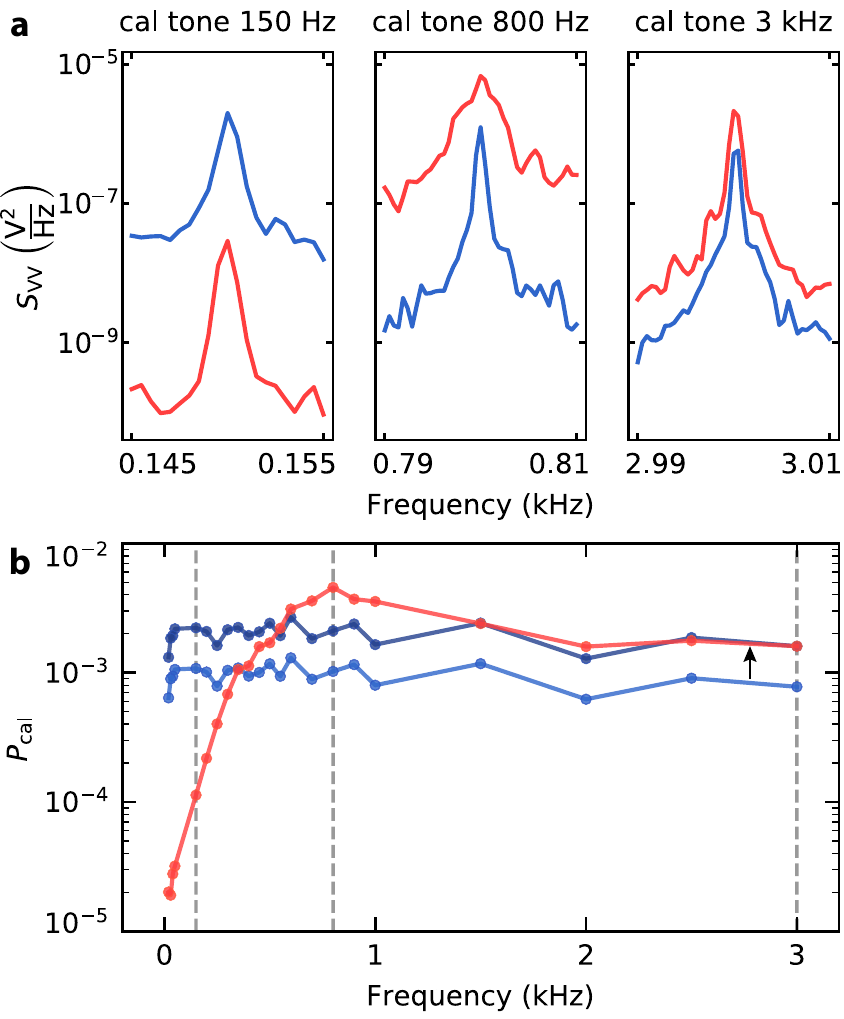}
    \caption{Calibration of the unlocked power spectral density. (a) Three measured homodyne power spectral density peaks, produced by an addition modulation tone applied to the local oscillator, for the unlocked (light blue) and locked (red) cavity. The calibration tones were provided at different frequencies inside and outside the feedback bandwidth. (b) The integrated power of the measured calibration tones $P_{\rm cal}$, using the same color coding. The dashed lines show the frequencies of the peaks in (a). The unlocked trace is corrected (dark blue) with an offset of the calibration factor.}
    \label{FigureSI3}
\end{figure}

The fluctuations of the unlocked cavity resonance frequency were greater than the region of linear response of the homodyne voltage, as shown in Fig.~\ref{Figure2}(b). As a result, the measured homodyne voltage did contain information about large frequency fluctuations. To compare the unlocked and locked noise spectra, separate calibration measurements had to be performed. With the Red Pitaya, an additional modulation was applied to the local oscillator at different frequencies for the cavity unlocked and locked. The amplitude of this tone caused a frequency fluctuation of the drive tone that was well within the linear regime of the locked homodyne response and resulted in a sharp peak in the measured power spectral density at the modulation frequency. For the locked cavity, when the frequency of this tone was within the feedback bandwidth, the feedback would create a control voltage to compensate for this artificially added noise. However, for frequencies greater than the feedback bandwidth, the feedback could not compensate this. Therefore, for the tones that were well outside the feedback bandwidth, the integrated area of the calibration tonne is equal for the locked and unlocked spectra. By comparing the integrated areas of these measured peaks, we found the correction factor to scale the unlocked power spectral density accurately. Figure~\ref{FigureSI3}(a) shows three added modulation peaks of the measured homodyne voltage, both for the unlocked and locked cavity: one within, one just outside, and one well outside the feedback bandwidth. Figure~\ref{FigureSI3}(b) shows the total measured power $P_{\rm cal}$ of the calibration tones applied. It can be seen that the feedback produces a high-pass filter behavior as expected. For the locked cavity, the powers of the calibration tones well outside the feedback bandwidth show a flat relation, indicating that the feedback did not reduce their noise power. The correction factor was then determined by the difference of the powers at the highest frequency, and the corrected unlocked trace was shown. Amending the unlocked data with this calibration enabled us to compare the unlocked and locked measurements fairly.

\bibliography{References.bib}

\end{document}